# Lo Gnomone Clementino: Astronomia Meridiana in Chiesa dal '700 ad oggi

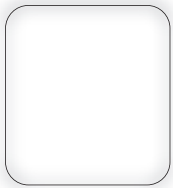

**Costantino Sigismondi**
Storia dell'Astronomia,
Sapienza, Università di
Roma, 6 ottobre 2010
sigismondi@icra.it

**Abstract**

The Clementine Gnomon is a giant pinhole dark camera dedicated to meridian solar astrometry operating in the Basilica of Santa Maria degli Angeli in Rome. Pope Clement XI ordered Francesco Bianchini (1662-1729) to build this instrument in 1701-1702. It renders solar images distortion-free, because the pinhole is optics-less. The azimut of the Clementine Gnomon has been referenced with respect to the celestial North pole, and it is 4'28.8"±0.6" Eastward. Also the local deviations from a perfect line are known with an accuracy better than 0.5 mm. Therefore the transit's times are systematically in delay with respect to the ephemerides. It is emphasized the opportunity of considering the Clementine Gnomon as introductory in modern astrometry besides its key role in the history of astronomy. Seeing effects on the solar image are studied using video. The need of a definitive solution in restoring the original pinhole is also shown.

**Sommario**

Ad una introduzione storica segue la descrizione della meridiana "ibrida" per Sole e stelle; la trovata della mira boreale per misurare la latitudine esatta; la deviazione dell'azimut della meridiana dal vero Nord celeste; le ragioni per cui questa deviazione è stata "importata" da un'altra meridiana usata come riferimento durante la costruzione. Per l'uso attuale dello Gnomone si riprende il concetto delle parti centesime dell'altezza e si descrive la tecnica dei transiti sulla linea meridiana e su una griglia di linee parallele ad essa per i video.

Come esempio di misure si presentano quella del solstizio estivo del 1703 scolpita sull'iscrizione marmorea nel presbiterio e realizzata usando anche i transiti meridiani diurni di Sirio; quella dell'equinozio autunnale del 2010 fatta da una foto digitale e quelle sulla turbolenza atmosferica sia ad alta che a bassa frequenza.

Infine si riprende il discorso del restauro del foro stenopeico e della finestrella per le osservazioni stellari.

**Introduzione**

Lo Gnomone Clementino è uno strumento meridiano a camera oscura con obiettivo a foro stenopeico: si trova nella Basilica di Santa Maria degli Angeli a Roma.

A 308 anni dalla sua inaugurazione (1702) questo strumento mantiene ancora intatto il suo fascino tanto per gli studiosi, quanto per gli appassionati di gnomonica o per visitatori occasionali. Anders Celsius (1701-1744), già professore di astronomia ad Uppsala, nei mesi trascorsi a Roma nel 1734, studiò a lungo e con ammirazione quella meridiana dichiarando perfino che questa compensava la banalità dei cortigiani di cui la città pullulava [3]. E c'è di che! Se la meridiana del Toscanelli fu la prima ad essere realizzata in Italia ed ha il foro più alto del Mondo, e quella di Cassini ebbe un tale successo scientifico che fu chiamato a Parigi dal Re Sole, la meridiana della "Certosa" di Roma è la più geniale creazione di questo genere, poiché funzionava con le stelle osservate di giorno per misurare direttamente con l'orologio a pendolo l'ascensione retta del Sole. Nessuna delle altre meridiane monumentali, costruite nel corso del secolo seguente, riprenderà questo design, limitandosi a segnare l'istante del mezzodì e la tangente dell'angolo zenitale del Sole.

Un mistero ne aumenta il fascino: già nella prima metà del settecento gli astronomi che visitarono la Basilica poterono constatare che la linea deviava lievemente dal Nord celeste: Celsius misurò 2' verso Est, mentre Ruggero Giuseppe Boscovich (1711-1787) nel 1750 misurò 4'30" [3] che è il valore misurabile anche oggi con gli strumenti più accurati. Le meridiane costruite dopo il 1750 non presentano più questo errore di azimut, ma la causa degli errori riscontrabili in tutte le grandi meridiane precedenti, inclusa la nostra, resta un problema ancora aperto nella storia dell'astronomia.

La ricognizione astrometrica dell'azimut della meridiana condotta nel 2006 ha consentito di validare le misure del Boscovich con maggiore pre-cisione: 4' 28.8"±0.6", così come è stato con le misure delle deviazioni locali dalla rettilineità prese con il laser tra il 2007 e il 2008.

L'attuale foro stenopeico è posticcio: ciò impedisce la ripetibilità delle misure a distanza di anni, aggiungendo ad un errore sistematico, oggi ben noto, una componente casuale che solo un restauro attento al valore scientifico e didattico dello strumento può eliminare.

Intendo una didattica di qualità metrologica, non quella che fa guardare i fenomeni senza misurarli.

Ripristinando il foro stenopeico esattamente a piombo sull'origine della meridiana, in posizionamento definitivo, si potranno calibrare col teodolite alcuni punti notevoli della meridiana, così da consentire le misure di astrometria solare con precisione millimetrica, corrispondente ai secondi d'arco.

L'introduzione delle videoriprese sincronizzate con il segnale orario RAI consente già di raggiungere precisioni assolute di 0.4 secondi nel tempo e 6" nelle posizioni angolari, ma il confronto su base annua non può essere fatto finché il foro non torni ad essere fisso. Le mura delle Terme di Diocleziano erano state ritenute le più adatte per questo strumento proprio per la loro stabilità che dopo 15 secoli avrebbe garantito al foro una posizione sempre fissa... almeno fino al 2500, che è la data dell'ultima orbita della Polare rappresentata sul pavimento della Basilica!

## Cenni storici

Lo Gnomone fu voluto e finanziato da papa Clemente XI Gianfrancesco Albani (1649-1721), tra i primissimi atti del suo pontificato iniziato il 23 novembre 1700, proprio nella stessa Basilica dove egli aveva celebrato la sua prima messa da sacerdote il 6 ottobre 1700 [1].

L'anno 1700 era il primo anno secolare non bisestile, a seguito della riforma del calendario di Gregorio XIII.

L'astronomo e archeologo Francesco Bianchini (1662-1729), veronese, allievo di Geminiano Montanari a Padova e scopritore di una cometa nel 1684, fu l'artefice di questo strumento di dimensioni colossali:

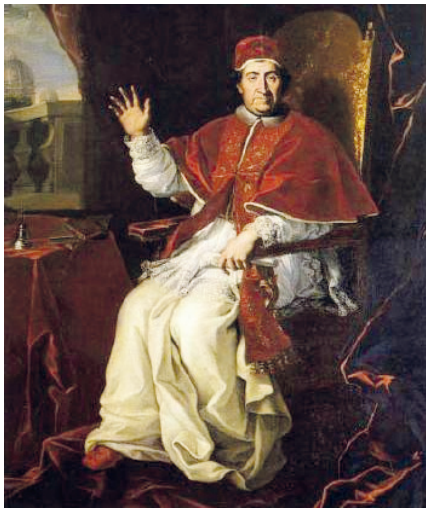 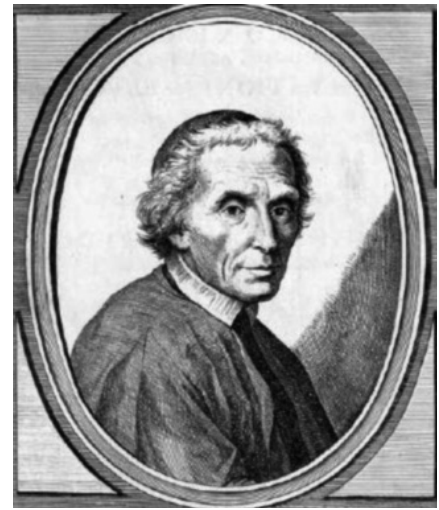

*Figura 1. Papa Clemente XI e Francesco Bianchini.*

20 metri di altezza e quasi 45 di lunghezza, realizzato con marmi preziosissimi ed inaugurato il 6 ottobre 1702 dal Papa.

Il 6 ottobre è anche la festa di san Bruno, fondatore dei Certosini, che tennero fino al 1887 la Basilica di Santa Maria degli Angeli, chiamata perciò anche "Certosa".

La scelta della Basilica, progettata nel 1564 da Michelangelo che trasformò in chiesa la grande aula con volte a crociera delle Terme di Diocleziano, risolveva anche il problema che Giandomenico Cassini (1625-1712) aveva riscontrato in S. Petronio a Bologna [5], dove la quota del foro stenopeico (del 1655) era variata leggermente in 40 anni a causa degli assestamenti dell'edificio. I medesimi problemi furono poi riscontrati dal gesuita Leonardo Ximenes (1716-1786) sullo gnomone (1475) di Paolo dal Pozzo Toscanelli (1397-1482) del duomo di Firenze [6].

Lo scopo dello Gnomone Clementino era quello di misurare i parametri dell'orbita terrestre: obliquità, eccentricità, e ancora la durata esatta dell'anno tropico, per chiudere la questione sulla precisione dei parametri della riforma gregoriana del calendario [3].

## Lo Gnomone Clementino: una meridiana ibrida per il Sole e per le stelle

Bianchini realizzò una meridiana sincronizzata con il tempo siderale.

Infatti alla meridiana di Santa Maria degli Angeli potevano osservarsi, anche simultaneamente, il transito meridiano del Sole e quello di alcune stelle fisse, aprendo la finestrella posta sopra il foro stenopeico. Bianchini nel suo libro *De Nummo et Gnomone Clementino* [1] riporta i dati dei transiti di Sirio e del Sole osservati nei giorni del solstizio d'estate del 1703.

I passaggi meridiani di una stella fissa consentivano di sincronizzare il pendolo di riferimento (fabbricato da Thuret a Parigi, il migliore del tempo) con la durata del giorno siderale pari ad 86164 secondi. Con quel pendolo si potevano contare le oscillazioni occorse tra un passaggio e l'altro della stella (tipicamente 86164±10) per calibrarne la durata media per ogni giorno, soggetta a variazioni con la temperatura.

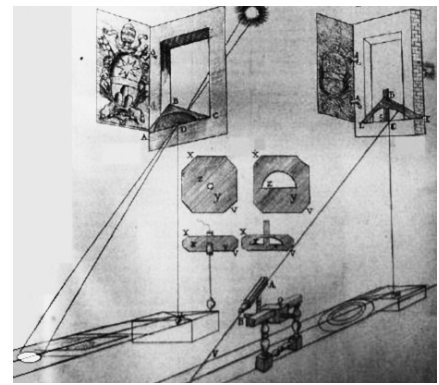

*Figura 2. Tratta dal testo di Bianchini: con un telescopio in asse con la meridiana, anche di giorno, si osserva il passaggio di una stella (destra), e il mezzogiorno locale (sinistra) con l'immagine stenopeica proiettata.*

In questo modo si poteva anche misurare direttamente la differenza in ascensione retta tra il Sole e la stella scelta come riferimento, rinviando alla precisione del catalogo stellare (Bianchini usò il catalogo del 1701 di Philippe de la Hire 1640-1718) quella sulla posizione del Sole.

L'osservazione diurna del passaggio meridiano delle stelle veniva fatta al telescopio osservando attraverso la finestrella aperta, non attraverso il foro stenopeico che è troppo piccolo. Il disegno di fig. 2 non lasci intendere che la luce delle stelle venisse proiettata sulla linea meridiana come accade col Sole: si tratta solo di un prolungamento geometrico. La misura delle ascensioni rette del Sole comparate con quelle delle stelle fisse consentì il calcolo degli istanti dei solstizi e degli equinozi pubblicati sulla tavola marmorea ora nel presbiterio della Basilica.

## La mira boreale per la Polare e la misura della latitudine

La determinazione della latitudine serviva a tracciare la retta equinoziale sul pavimento e si faceva con i transiti al meridiano della stella Polare.

La linea meridiana boreale si sviluppa sul pavimento (quasi) parallelamente a quella australe, ma solo per i primi 6 metri, e taglia a metà le ellissi descritte dalla proiezione dell'orbita giornaliera della stella Polare attorno al polo Nord celeste, con funzione didattica e decorativa, calcolate per tutti gli anni giubilari dal 1700 al 2500.

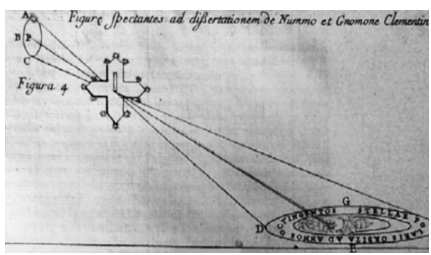

*Figura 3.* La croce-mira boreale attraverso cui si osservarono i transiti superiore (A) ed inferiore (C) della Polare, con l'asse del telescopio allineato con la linea meridiana. P è la posizione del polo Nord celeste. Il finestrone in corrispondenza della croce-mira veniva aperto quando Bianchini osservava la Polare dall'interno della Basilica.

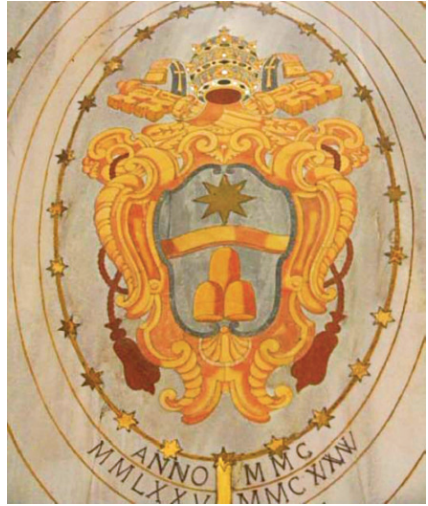

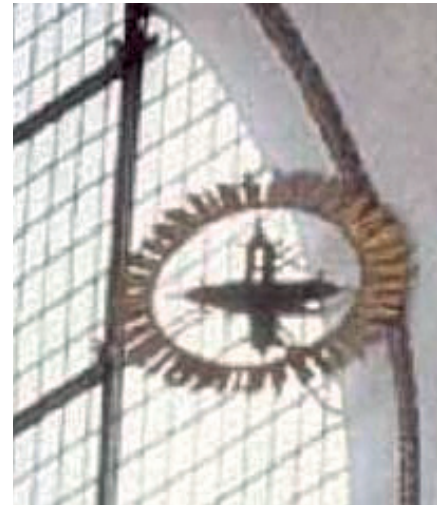

*Figura 4.* Proiezione del polo Nord celeste al centro delle orbite giubilari, e croce-mira boreale per i transiti superiori ed inferiori della stella Polare.

Bianchini, al telescopio dall'interno della Basilica con l'asse allineato sulla linea boreale, osservò i transiti superiore ed inferiore della Polare nelle notti tra l'1 e l'8 gennaio 1701, quando la Polare culminava superiormente alle 6 di sera ed inferiormente alle 6 del mattino, sempre con il cielo buio.

La posizione media P in fig. 3 corrisponde al polo Nord celeste apparente.

Bianchini corresse questo dato per la rifrazione atmosferica ed ottenne 41°54'27" (fece scrivere 30" sul marmo). Oggi con il GPS misuriamo per il foro stenopeico 41° 54'11.2" avendo come ellissoide di riferimento il WGS84 oppure 41° 54' 08.86" con il ROMA40 che è il più adatto per le misure a Roma.

A causa dell'aberrazione della luce, scoperta da James Bradley nel 1727, anche la posizione media P della Polare compie un orbita annuale attorno al vero polo celeste con un semiasse maggiore pari a 20.45 secondi d'arco, e P ai primi di gennaio si trovava proprio al massimo spostamento verso Nord (+20.2"). La nutazione dell'asse terrestre, altra scoperta di Bradley del 1737, aveva nello stesso periodo una componente in declinazione pari a −4.8", così che P ai primi di gennaio 1701 appariva 15.4" più a Nord del polo Nord celeste vero.

Sottraendo le componenti dovute all'abberrazione e alla nutazione al dato di Bianchini si troverebbe [9] per la latitudine del foro stenopeico 41° 54'11.6": un valore entro soli 3" dal dato ROMA40.

A meno di una coincidenza fortunata di combinazioni di errori sperimentali più grandi, questo fatto indica che Bianchini era in grado di determinare la proiezione di P sul pavimento della Basilica a partire dalla croce-mira a 26 metri di altezza con precisione millimetrica.

Le orbite della stella polare in tutti gli anni giubilari sono il frutto del calcolo della precessione degli equinozi, che modifica progressivamente la longitudine eclitticale di tutte le stelle, lasciandone invariata la latitudine.

Perciò la stella Polare compie un'orbita attorno al polo Nord dell'eclittica che la poterà attorno al 2100 alla minore distanza dal polo Nord celeste, pari a 28'.

La stella al centro dello stemma di papa Clemente XI corrisponde al polo Nord celeste, nella foto 4 a è presente l'orbita corrispondente al 2100 e tratti delle orbite tra il 2000 ed il 2200. Oggi (2010) la Polare si trova a 41' dal Nord celeste.

## La deviazione della meridiana australe

Da misure fatte al teodolite referenziato con la stella Polare [2] sappiamo che la meridiana australe devia dalla direzione esatta del Nord di 4'28".8±0.6" verso Est [9], che corrispondono a circa 6 cm alla distanza di 45 metri dal piede della verticale del foro stenopeico. La ragione di questa discrepanza non è stata ancora spiegata. Si sa che è un difetto presente anche su strumenti analoghi coevi e precedenti e tende a sparire alla fine del secolo XVIII [8]. Per motivi non chiari anche la linea meridiana boreale non è esattamente parallela a quella australe, ma ne diverge di 2 cm verso Ovest a circa 35 metri di distanza, in corrispondenza del piede della verticale della croce-mira dello gnomone boreale.

Difetti di costruzione e successivi interventi di restauro un po' approssimativi hanno determinato deviazioni locali dalla rettilineità della linea australe di qualche mm (punti rossi sulla linea reale in fig. 5), ma la deviazione iniziale dal Nord ed un ampio seno sono rimasti tali sin dall'inizio, cioè prima del rifacimento del pavimento da parte di Luigi Vanvitelli in occasione del giubileo del 1750.

La prova che la deviazione è originale la ho trovata analizzando le misure combinate dei transiti Sirio-Sole dell'estate 1703 [1]. Sirio allora passava attorno al numero 161 (cfr. fig. 6), mentre il Sole era sul Cancro vicino ai numeri 33-34, e per effetto della deviazione della linea gli istanti di passaggio di Sirio hanno un ritardo relativo a quelli del Sole di circa 11 secondi [9], rispetto a quanto si sarebbe potuto osservare da una meridiana perfettamente allineata con il Nord, e quindi calcolare con un programma di effemeridi [7, 11].

Oggi se potessimo osservare Sirio con un telescopio attraverso la finestrella della meridiana, la proiezione dell'asse ottico non ca-drebbe più sul numero 161. La precessione degli equinozi ha modificato di due intere tacche la sua posizione sulla linea, ed il moto proprio della stella di circa 1.2 secondi d'arco all'anno verso Sud di altre 0.65 tacche, col risultato che oggi Sirio si potrebbe osservare quasi dal numero 164.

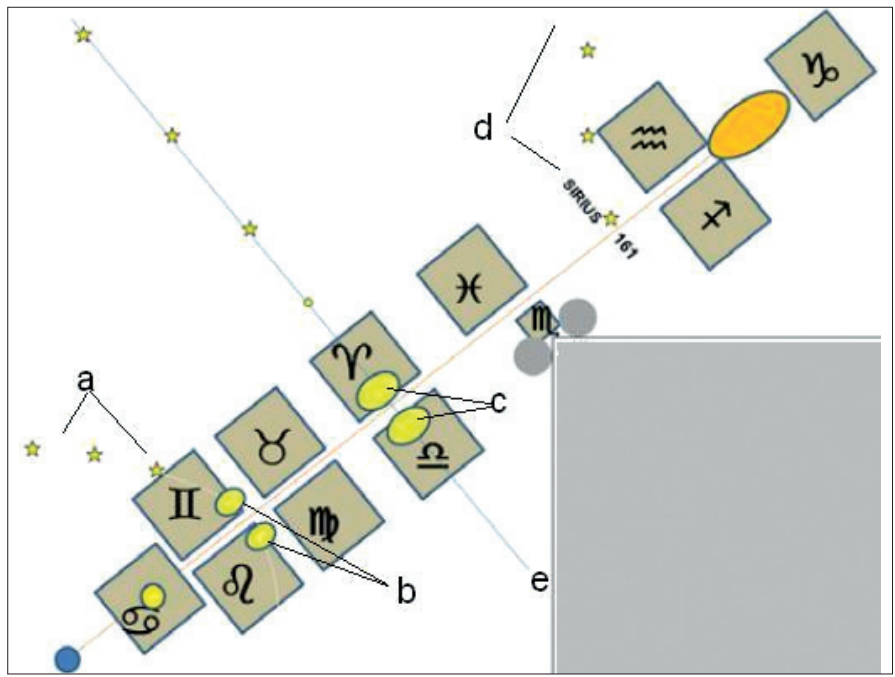

*Figura 7. Il Capricorno in Basilica e quello di Bayer.*

Nella pianta della meridiana lo Scorpione è più piccolo degli altri poiché si trova presso una colonna delle antiche Terme, un monolite di granito alto 14 metri, che impedisce l'osservazione del Sole già sei minuti dopo il transito tra Novembre e Febbraio.

La linea perpendicolare alla meridiana è l'equatore celeste (e), sull'Ariete e la Bilancia c'è il cronometro degli equinozi, rappresentato dalla coppia di immagini (c) a ca-vallo della Meridiana. Un'altra coppia di immagini solari di bronzo è in corrispondenza del punto di transito del Sole il 20 agosto: ricorda la visita estemporanea di Clemente XI nel 1702 in occasione della festa di S. Bernardo (b) nella vicina chiesa "alle Terme", preceduta da un arco iperbolico di stelle (a).

L'altro arco di iperbole con la concavità rivolta verso il foro -non rappresentato nella figura- è per Arturo, e quello che interseca la Linea al n. 161 è per Sirio (d), con la concavità opposta.

La parete ovest del vano esterno (fig. 14) dov'è ricavato il foro stenopeico fa ombra sul foro stesso già 20 minuti dopo il transito meridiano.

Per queste due ragioni strutturali la me-ridiana non poté essere orientata con la bisettrice dei punti dove il Sole attraversa un cerchio di altezza 1 o 2 ore prima e dopo il meridiano, ma fu allineata secondo una meridiana ausiliaria che Giacomo Filippo Maraldi (1665-1729) aveva realizzato nel Palazzo Venezia [3].

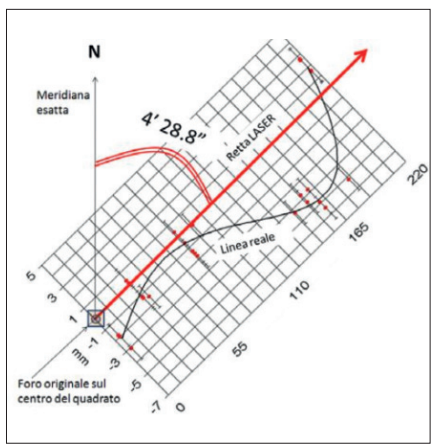

*Figura 5. Deviazione generale verso Est della Linea Clementina in funzione delle parti centesime (fino a 220) dell'altezza del foro stenopeico, e deviazioni locali rispetto ad una retta laser.*

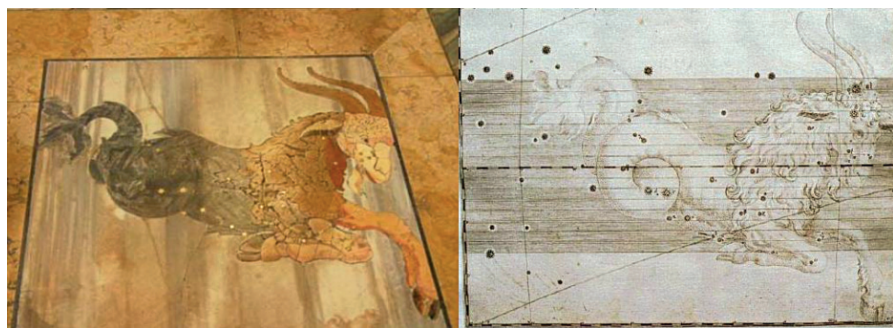

La deviazione verso Est fu dunque importata da un'altra meridiana, mediante segnali ottici per sincronizzare il mezzodì.

I segni zodiacali furono realizzati con la tecnica delle tarsie marmoree, seguendo i modelli delle costellazioni disegnate da Johannes Bayer nella sua Uranometria (1603). Il confronto tra il Capricorno e la carta stellare è preciso nei dettagli. La linea si estende fino a 220.6 (44 m 90) per includere la massima elongazione del bordo meridionale del Sole al solstizio d'inverno al principio del XVIII secolo, quando l'obliquità media dell'eclittica valeva 23°28'40".

### Principi di funzionamento dello Gnomone

Il Sole transita sull'orizzonte Sud nel punto più alto del suo corso giornaliero, e, da un giorno all'altro, la sua altezza sull'orizzonte romano della "Madonna degli Angeli" varia da un minimo di 24°41' al solstizio invernale ad un massimo di 71°32' a quello estivo.

Nell'immagine in fig. 8 ho usato un colore più scuro per l'immagine ed il Sole invernale, che è anche più grande, mentre il colore è più chiaro nel caso estivo.

Le ragioni sono due: d'inverno la massa d'aria attraversata dai raggi solari nell'atmosfera terrestre è oltre il doppio di quella d'estate, perciò l'intensità luminosa è minore; per effetto dell'inclinazione l'area dell'immagine del Sole diventa circa 12 volte più grande, con una conseguente diminuzione di intensità di 12 volte.

### Parti centesime dell'altezza: i numeri da 38 a 220

Sulla linea meridiana si sviluppano due numerazioni. La prima è fatta di numeri equispaziati, ad una distanza di 203.44 mm gli uni dagli altri. Ogni intervallo corrisponde ad 1/100 dell'altezza del foro sul pavimento.

Ogni numero sulla linea corrisponde anche a 100 volte il valore della tangente dell'angolo zenitale (indicato nell'altra numerazione) che è l'angolo tra il centro del foro, il punto sulla linea e lo zenith.

Se z è la distanza angolare dallo zenit del Sole, valgono le seguenti relazioni: Parte Centesima=100·tan(z) e quella inversa z=arctan(Parte Centesima/100).

Ad esempio il Sole in eclissi il giorno 29 marzo 2006 è passato vicino al numero 78, a 38 gradi dallo zenit, poiché 78≈100·tan(38°).

### Misure del tempo di un transito meridiano

Lo strumento è concepito principalmente per misurare dei tempi, perciò bisogna usarlo sempre con un orologio di riferimento. Bianchini aveva i tempi di riferimento da un pendolo, i cui rintocchi venivano confrontati con i transiti sulla meridiana delle 22 stelle di riferimento, i cui nomi e ascensioni rette sono riportati sul marmo in corrispondenza del loro angolo zenitale nel 1701.

Noi possiamo supplire alla mancanza dell'osservazione delle stelle con un orologio radio controllato.

Per misurare gli istanti di passaggio al meridiano occorrono i due tempi di primo e secondo contatto, come indicato nello schema di fig. 9b. L'immagine del Sole viaggia da destra a sinistra, il Sud è in alto ed il tempo di transito si calcola con la formula tc=(t1+t2)/2.

### Il cronometro degli equinozi

Bianchini ha concepito questo strumento per consentire una valutazione rapida dell'istante in cui il centro del Sole si trova sull'equatore celeste. E' costituito da due ellissi uguali poste a destra e a sinistra della linea meridiana; sugli assi maggiori di queste ellissi ci sono due regoli graduati che servivano a leggere quante ore erano passate o mancavano all'equinozio. Sull'asse minore passa la linea dell'equatore celeste.

Oggi la scala graduata è stata obliterata dal calpestio, ma è possibile ugualmente calcolare l'istante dell'equinozio come di seguito.

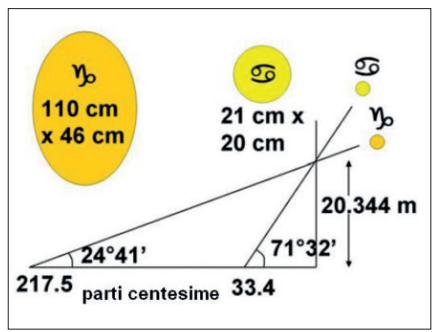

*Figura 8.* Solstizi e altezza meridiana del Sole a Roma.

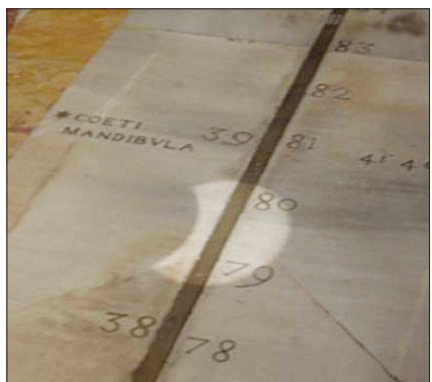

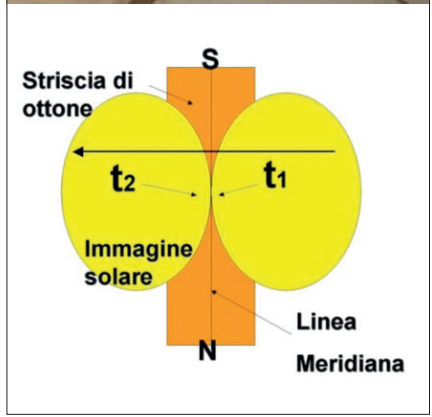

*Figura 9.* a. Eclissi del 29.3.2006; b: Schema dei transiti.

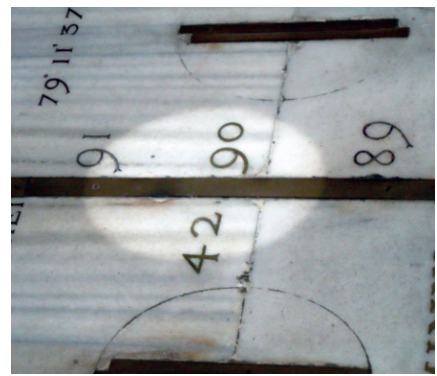

*Figura 10.* Equinozio autunnale 2010, foto di Enrico Giuliani.

Analizzando la foto digitale in fig.10 del 23 settembre 2010 alle 13:02 il centro del Sole ha superato di 157±1 pixel la linea equatoriale (quasi verticale nella foto a destra di 42-90), il diametro meridiano esclusa la penombra di 14±2 pixel vale 608±2 pixel e corrisponde ripettivamente a 31 ore nel moto in declinazione del Sole e a 324 mm sul pavimento [4], per cui l'equinozio è avvenuto 8h00±3min prima, mentre risulta 7h43m prima secondo le effemeridi IMCCE [www.imcce.fr]. Questi 17±3 minuti di anticipo mostrano che l'attuale foro stenopeico provvisorio è spostato verso Nord di circa 3 mm. La scritta (Term)inus Pasch(ae) ricorda che l'equinozio di primavera è il limite (Terminus) inferiore per la data della Pasqua, mentre a sinistra è la stella (Orioni)s Balthei (Prim)a con la sua ascensione retta 79°11'37".

Nella figura 11 si rappresenta schematicamente il metodo per proiettare le stelle sulla linea meridiana. Le loro declinazioni sono riportate sul cerchio centrato nel foro e tangente al pavimento.

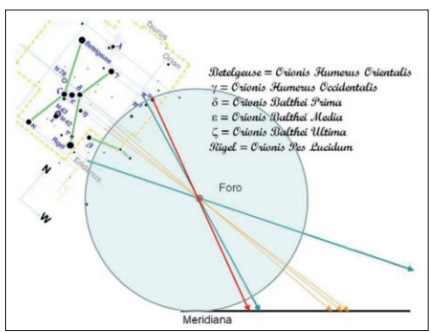

*Figura 11.* La proiezione delle stelle di Orione; le stelle della cintura (Prima, Media e Ultima) sono presso l'equatore celeste ed il cronometro degli equinozi. Immagine: Magdalena J. Bedynski e Monica Nastasi.

### La misura del solstizio estivo del 1703

Sirio era la stella di riferimento per i transiti meridiani durante il solstizio estivo. Il passaggio al meridiano di Sirio veniva osservato sistematicamente in ritardo a causa della deviazione verso Est della linea Clementina. Per questa ragione Bianchini riteneva che il Sole raggiungesse la stessa ascensione retta di Sirio, ed anche quella del solstizio, in ritardo rispetto al vero.

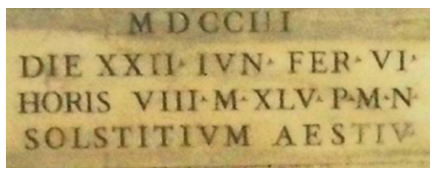

*Figura 12.* Solstizio estivo venerdì 22.6.1703 alle 8:45 dopo la mezzanotte locale "P(ost)M(edia)N(octe)", che era stata alle 23:11 UT; il solstizio fu calcolato alle 7:56 UT, 33 minuti di ritardo sulle effemeridi IMCCE che danno 7:23 UT. Lastra di marmo nel presbiterio.

Il Sole percorre in cielo circa 4 minuti di ascensione retta ogni 24 ore, per cui la differenza sistematica di 11 s di ascensione retta corrisponde a 66 minuti di tempo, di questi solo 33 minuti risultano dal calcolo del solstizio estivo del 1703: molto probabilmente Bianchini usò anche altre stelle più a Nord di Sirio, e quindi con meno ritardo sistematico. Al solstizio invernale la situazione si inverte: le stelle di riferimento sono tutte più a Nord del Sole e quindi passano in anticipo sulla meridiana così come anticipata è la stima di quel solstizio [9]. Per gli equinozi le stelle di riferimento sono sia a Nord che a Sud, quindi gli errori tendono a compensarsi. Il massimo errore commesso da Bianchini per il 1703 è proprio quello di 33 minuti al solstizio estivo.

### Ottica atmosferica: turbolenza e seeing

L'immagine solare sul pavimento della Basilica appare in continua vibrazione, ad alta frequenza.

Il foro stenopeico all'esterno si trova su un piano di mattoni che al mezzodì è scaldato direttamente dal Sole: l'aria calda si solleva generando turbolenza. Il flusso d'aria che attraversa il foro (senza ottica, perciò aperto) per le differenze di temperatura tra interno ed esterno, costituisce un'altra sorgente di turbolenza.

Le bolle d'aria sono di dimensione maggiore del foro per cui tutta l'immagine del Sole viene influenzata contemporaneamente.

L'effetto della turbolenza locale e di quella propria dell'atmosfera che si genera tra strati a differenti velocità costituisce il *seeing* astronomico. Sull'immagine stenopeica si possono studiare le componenti del *seeing* ad alta frequenza (*blurring*) e a bassa frequenza (*image motion*), mediante delle riprese video fino a 60 fotogrammi per secondo.

Vengono predisposti dei fogli con N righe parallele [10] equispaziate, disposte in modo che quella centrale coincida con la linea meridiana (fig.13).

Il passaggio dell'immagine del Sole su questo foglio viene filmato a 60 fps e poi riesamianto un fotogramma alla volta. Su una tabella si dispongono i tempi in cui uno stesso lembo del Sole tocca in sequenza tutte le righe.

Le differenze tra questi tempi di contatto, se l'atmosfera fosse calma, sarebbero tutte uguali, ma a causa della componente ad alta frequenza del seeing c'è una dispersione attorno alla media di ±0.4 secondi.

Nello stesso intervallo di tempo l'immagine del Sole percorrerebbe 6" in cielo: questo valore angolare corrisponde esattamente con il limite di diffrazione in luce visibile da un'apertura di 15 mm. Questo fatto significa che la risoluzione spaziale e temporale delle nostre immagini stenopeiche è limitata dalla diffrazione.

Poiché il limite di diffrazione è inversamente proporzionale al diametro, nei telescopi con obbiettivi più grandi, quasi sempre il seeing è più grande della diffrazione. Applicando questo stesso metodo al telescopio solare dell'IRSOL [www.irsol.ch] di 45 cm di diametro (limite teorico di diffrazione 0.2") abbiamo misurato talvolta valori eccellenti di seeing diurno pari a 0.6" [12] con immagini molto ben definite.

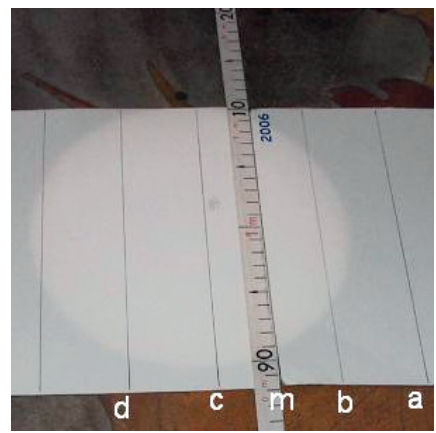

*Figura 13.* Passaggio dell'immagine solare su linee parallele alla meridiana con macchia solare: 1.7.2006.

Le linee parallele sono state concepite anche per misurare l'istante del passaggio al meridiano con la massima precisione possibile. Se il primo bordo del Sole tocca le linee a, b, m (meridiana), c, d nei tempi t1a, t1b, t1m, t1c, t1d, ed il secondo bordo nei tempi t2a, t2b, t2m, t2c, t2d, abbiamo 5 determinazioni indipendenti del tempo di passaggio sul meridiano:

tm1=(tm1+tm2)/2;
tm2=(ta1+td2)/2;
tm3=(tb1+tc2)/2;
tm4=(ta2+td1)/2;
tm5=(tb2+tc1)/2;

Estendendo questo procedimento ad una griglia di N linee, si ottengono N determinazioni indipendenti dell'istante del mezzodì. Poiché un transito di tutta l'immagine del Sole dura circa 2 minuti le coppie di tempi che producono tm1, tm2 etc... sono separate da 2 minuti in su. La dispersione della media degli N tempi tm1...tmN non si riduce di un fattore $1/\sqrt{N}$ (secondo il teorema sulla varianza della media di N dati omogenei e gaussiani) ma resta a ±0.4 secondi.

Questo è un indice che l'immagine del Sole subisce anche un *image motion* a bassa frequenza su tempi scala superiori ai 2 minuti, come quelli che intercorrono tra ciascuna delle N coppie con media tm, che noi osserviamo sempre combinata con la diffrazione del foro stenopeico che ne limita l'ampiezza minima osservabile a 6".

### Ripristino del foro originale di Bianchini

Attualmente il foro è una maschera di plastica con un'apertura tonda di 1.5 cm di diametro applicata sopra un foro preesistente a forma di fagiolo [4] ritagliato co-sì nel metallo. Ogni forma diversa dal cerchio produce un'immagine stenopeica con penombra asimmetrica, che peggiora la qualità delle misure alla meridiana.

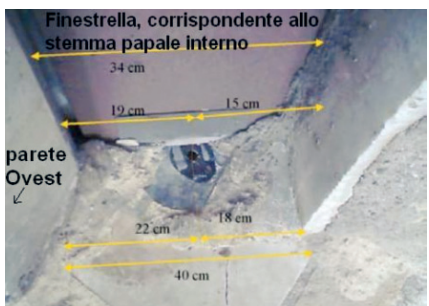

*Figura 14.* Il foro stenopeico all'esterno, interno in fig. 2.

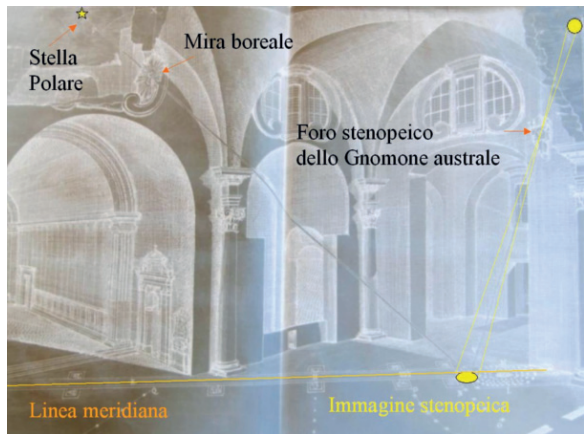

*Figura 15.* Schema della meridiana australe e della mira boreale in Santa Maria degli Angeli, adattato dal testo di Bianchini [1].

Col filo a piombo, già previsto dal Bianchini, basterebbe poco per ricollimare la struttura, facendo un foro tondo di 20.3 mm esattamente sulla verticale del centro del quadrato in marmo sotto il foro stenopeico. Realizzandolo in invar verniciato di bianco al biossido di titanio si ridurrebbe la turbolenza locale e l'espansione termica, massima al mezzodì. Un restauro anche della finestrella, con possibilità di riaprirla, consentirebbe le osservazioni stellari diurne alla meridiana, e sarebbe di grande richiamo turistico e profondo valore culturale.

## Bibliografia